**The physical basis of natural units and truly fundamental constants**

Leonardo Hsu[a]
Department of Postsecondary Teaching and Learning, University of Minnesota, Minneapolis, Minnesota 55455, USA

Jong-Ping Hsu[a]
Department of Physics, University of Massachusetts-Dartmouth, North Dartmouth, Massachusetts 02747, USA

Abstract:

The natural unit system, in which the value of fundamental constants such as $c$ and $\hbar$ are set equal to one and all quantities are expressed in terms of a single unit, is usually introduced as a calculational convenience. However, we demonstrate that this system of natural units has a physical justification as well. We discuss and review the natural units, including definitions for each of the seven base units in the International System of Units (SI) in terms of a single unit. We also review the fundamental constants, which can be classified as units-dependent or units-independent. Units-independent constants, whose values are not determined by human conventions of units, may be interpreted as inherent constants of nature.

[a] e-mail: lhsu@umn.edu, jhsu@umassd.edu

## I. Introduction

In any field of study, operational definitions are critical for minimizing misunderstandings and providing for efficient communication within its community of practitioners. In a quantitative experimental science such as physics, a system of units is one form in which operational definitions are realized.

"Natural units" is a system of units in which the vacuum speed of light $c$ and Planck's constant $\hbar$ are dimensionless with unit magnitude. All physical quantities are then expressed in terms of a power of a single unit, usually mass or energy.[1] One useful feature of natural units is that the equations expressing physical laws are then simpler to write down and have fewer constants obscuring the essential physics they embody. This unit system is learned by the vast majority of physics students at some point in their advanced undergraduate courses and it is widely used by theoretical particle physicists.

We should state clearly however, that our aim is not to advocate for a reduction in the number of units nor to propose the wider general use of the natural unit system. There are numerous practical and pedagogical reasons why the SI system, CGS system, or other unit systems are better suited for most practicing physicists and physics students, including the ability to check one's work by verifying the units of the answer, the ability to perform dimensional analyses to infer functional dependences, and simple familiarity and comfort with a more standard unit system when trying to learn already challenging material. Instead, our aim is to investigate whether this unusual unit system is based in the physical laws of our universe or whether setting $c$ and $\hbar$ to one is simply a calculational convenience.

In most scientific work, quantities are expressed in some system of units, often the International System of Units (SI), which is the modern metric system of measurement. In the SI, there are seven base units, the meter, second, kilogram, ampere, kelvin, mole, and candela, and many more derived units, which are products of powers of the base units.[2] It is important to note that the definitions of these base units are interdependent, for example, the definition of the ampere incorporates those of the meter, kilogram, and second, even though the base quantities corresponding to those base units (length, time, mass, electric current, thermodynamic temperature, amount of substance, and luminous intensity, respectively) are conventionally regarded as independent. Other systems of units have a similar structure. The choice of the seven base units in SI is somewhat arbitrary, but was set by the General Conference on Weights and Measures (CGPM) based on a number of factors, including history, practicality, accuracy, and reproducibility. More importantly, each of these base units is defined in terms of some physical property or artifact, linking it to reality and enabling the comparison of any quantity to a physical standard.

In order for natural units to be said to have a physical basis, in effect reducing the seven base units in SI to a single unit, one must be able to define operationally each of the base SI units in terms of a single unit in a physically meaningful way. For example, one could arbitrarily decide that 1 meter is equivalent to 5 ampere, but such an equivalence would not be physically meaningful because no law of physics or physical phenomenon supports such an equivalence. On the other hand, saying that 1 meter is equivalent to 1/299 792 458 second **is** physically meaningful because such an equivalence is rooted in a physical law, namely that electromagnetic waves propagate in accordance with the equation $r^2 - c^2 t^2 = 0$ where $c$ is the universal speed of light [3,4] and the maximum possible speed in our universe, 299 792 458 m/s in all inertial frames. Indeed, in the SI system, the meter has been defined since 1983 as "the length of the path travelled by light in vacuum during a time interval of 1/299 792 458 of a second"[2] and thus although the meter and the second are still considered dimensionally independent in the SI system, the definition of the meter is now dependent on the second.

In this review paper, we show that there is indeed a physical basis for the natural unit system, developing definitions and physical meaningful equivalences for all the base SI units in terms of a single unit. Indeed, new definitions for the base SI units of a similar nature have been proposed in the literature and will be considered and perhaps adopted by the CGPM, possibly as early as 2011.[5] In the following section, we use special relativity and the quantum theory to show how the dimensions of mass, length, and time can be redefined in terms of a single unit. In section III, we derive definitions for the remaining four base SI units in terms of our single unit, and briefly comment on units for other common physical quantities. In section IV, we compare the natural unit system described here with another "natural unit" system, Planck's natural units. We also briefly discuss the changes others have proposed in the definitions of the SI units and their similarity to the definitions we present here. Finally, we investigate the fundamental constants using natural units and suggest that only those constants whose values are units-independent should be interpreted as inherent and true constants of nature.

## II. Time, length, and mass

### A. Time and length

Prior to the twentieth century, space and time seemed to be completely different entities and thus two different units, the meter and the second, were invented to quantify them. From a modern viewpoint however, the theory of special relativity implies that space and time are not independent and separate, but parts of a four-dimensional spacetime.[6] Interestingly, this view is consistent with the definition of the SI unit of length, meter, which is given in terms of that of the unit of time, second, although that particular definition is the result of a number of issues of accuracy and practicality, rather than a change in our fundamental view of the universe.[2]

In our physical laws, four coordinates–three spatial and one temporal–are necessary to specify the spacetime location of observable events. Although there is nothing wrong with using different units to express these four quantities it is logically and mathematically simpler to use the same single unit, either meter or second. As an analogy, there is nothing wrong with expressing north-south distances in miles and east-west distances in meters. Doing so, however, introduces an extra conversion constant into physical equations that is clearly artificial as the spatial distance $s$ between any two points on a two-dimensional surface would then be given by $s = (x^2 + k^2y^2)^{1/2}$ where $x$ is the east-west separation between two points, $y$ is the north-south separation, and $k$ is a conversion factor 1609 meter/mile (this example comes from the Parable of the Surveyors in Taylor and Wheeler's Spacetime Physics (Ref. 7, p. 1)). Expressing both north-south and east-west distances in the same unit enables the simpler expression of the distance as $s = (x^2 + y^2)^{1/2}$. The mathematical form of physical laws is much simpler when distances in both directions are expressed using the same unit. The same is true for spacetime intervals. The square of the spacetime interval $s$ in special relativity is $s^2 = r^2 – c^2t^2$ if one measures spatial intervals in meters and time intervals in seconds, but is simply $s^2 = r^2 – t^2$ if one measures both spatial and time intervals using the same unit.[7,8]

In order to create a physically meaningful equivalence between the meter and the second, there must exist a physical law or phenomenon that connects both length and time in an invariant way. The law for the propagation of light $r^2 – c^2t^2 = 0$ is just such a phenomenon and the speed of light $c$, which is also the maximum speed of a particle with a non-negative, non-imaginary mass provides a convenient conversion factor between the two units. From a purely theoretical point of view, either the meter or the second could be used as the fundamental unit, with other units defined in terms of the fundamental one. For practical reasons, in the SI unit system, the definition of the meter was made dependent on the second rather than having the definition of the second be dependent on the meter. The modern (1983) definition of the meter is "The meter is the length of the path travelled by light in vacuum during a time interval of 1/299 792 458 of a second." [2]

One could then say that one meter is equivalent to 1/299 792 458 second. One of the consequences of such a definition of the meter is that the speed of light $c$ now has an exact specified value, equal to 299 792 458 meter/second, and plays the role of a conversion factor between meters and seconds.[7] In other words, $c$ now has the exact value 1 if lengths and time intervals are measured using the same units and, like any other velocity, is dimensionless.

### B. Time and mass

At present, the definition of the kilogram, the SI unit of mass, is based on the international prototype of the kilogram, an object made of a platinum-iridium alloy stored at the International Bureau of Weights and Measures in Sèvres, France. This definition, which reads "The kilogram is the unit of mass; it is equal to the mass of the international prototype of the kilogram"[2] is independent of all other units. However, just as the four-dimensional symmetry of our universe gives us a physical basis to unify the definitions of length and time, the quantum nature of our universe provides a physical basis for the unification of the unit of mass with that of time and length. Because of the precedent set by the SI system in defining the meter in terms of the second, we will continue to use the second as our fundamental unit in all further discussions, although all of the proposed definitions could easily be re-cast to use, say the meter or the kilogram as our fundamental unit.

As before, in order to develop a physically meaningful equivalence between the unit of mass and the unit of time, there must be a physical law or phenomenon linking both types of quantities. The equivalence of mass and energy in relativity theory provides just such a phenomenon. As we saw in the previous section, when lengths and time intervals are expressed using the same single unit, the speed of light $c$ is dimensionless with the value of unity and thus the equation $E = mc^2$ can be written $E = m$. In the quantum theory, the energy of a photon is related to its frequency through $E = h\nu$. Putting these two equations together gives $m = h\nu$, a relationship linking the mass of a particle to the frequency of a photon (or total frequency of a collection of photons) that has the same energy (mass) as the particle.

Just as the definition of the meter resulted in the setting of the value of $c$ to unity, the kilogram can be defined in such a way as to set the value of $h$ to unity, making $h$ a conversion factor between kilogram and second. Using the equivalence of 1 meter to 1/299 792 458 second, we get $h = 6.626\ 069\ 3 \times 10^{-34}$ J•s = $6.626\ 069\ 3 \times 10^{-34}$ kg•m$^2$/s = $6.626\ 069\ 3 \times 10^{-34}$ $(1/299\ 792\ 458)^2$ kg•s = 1, so that 1 kg is equivalent to $[(299\ 792\ 458)^2/662\ 606\ 93] \times 10^{41}$ s$^{-1}$. Expressing masses and time intervals using the same units makes the quantities of action and angular momentum dimensionless.

Following this line of thought, one possible definition of the kilogram that has been proposed in the literature is: "The kilogram is the mass of a body whose equivalent energy is equal to that of a number of photons whose frequencies sum to exactly $[(299\ 792\ 458)^2/662\ 606\ 93] \times 10^{41}$ hertz."[5] As with the definition of the meter, one of the consequences of such a definition of the kilogram is that the value of Planck's constant $h$ now has an exact specified value, equal to $6.626\ 069\ 3 \times 10^{-34}$ kg•s (or 1, if masses and time intervals are measured using the same units).

In natural units, it is more usual to set $\hbar = 1$ rather than $h = 1$, so that 1 kg is equivalent to $2\pi(299\ 792\ 458)^2/662\ 606\ 93] \times 10^{41}$ s$^{-1}$ since $E = \hbar\omega$. In this case, one could define the kilogram as "The kilogram is the mass of a body whose equivalent energy is equal to that of a number of photons whose **angular** frequencies sum to exactly $[2\pi(299\ 792\ 458)^2/662\ 606\ 93] \times 10^{41}\ s^{-1}$." For the remainder of this paper, we will use this alternate definition since it is more consistent with the natural units in use today. It is important to note that although the numerical value of the equivalence between the kilogram and inverse second is a matter of human convention and not unique, the main result, that there is a physical basis for expressing masses and time intervals using the same unit (albeit different powers of that single unit), still stands.

We now see that expressing distances, time intervals, and masses in terms of the same unit (by setting $c$ and $\hbar$ to 1) is not merely an artificial choice made purely for purposes of simplifying mathematical calculations. Instead, it has a physical basis in that in our universe, the three dimensions of length, time, and mass are all related in a fundamental way through the four-dimensional symmetry of spacetime and the quantum theory or their union, relativistic quantum mechanics.

## III. Other SI base units

We now develop corresponding definitions and equivalences for the other base SI units. The classification of the ampere, kelvin, candela, and mole as "base" units is a historical one and although in SI, they are dimensionally independent, that does not imply that those units are truly independent of the meter, second, and kilogram.[9] Indeed, the definitions of the ampere, mole, and candela are already dependent on the definitions of the meter, second, and kilogram.

### A. Temperature

The present definition of the SI unit of temperature is "The kelvin, unit of thermodynamic temperature, is the fraction 1/273.16 of the thermodynamic temperature of the triple point of water."[3]

However, just as physical laws relate space and time in special relativity, and mass and frequency in quantum mechanics, the kinetic theory of gases implies that temperature is not an independent characteristic of a system, but simply one type of energy scale that is closely related to the kinetic energy of the particles that make up the system. For example, the temperature of a monatomic ideal gas is related to the average kinetic energy of its individual gas molecules through $\overline{E} = (3/2)kT$.

To establish a physically meaningful relationship between the kelvin and other units, we note that the Boltzmann constant $k$ already plays the role of a conversion factor between two energy scales of a system, a microscopic scale involving the average kinetic energy of the particles and a macroscopic scale known as temperature. In a sense, the Boltzmann constant is similar to Joule's constant, which was used to relate mechanical energy (in joules) and heat energy (in calories). When it was realized that both were different aspects of the same type of quantity (energy), Joule's constant 4.184 J/cal became merely another conversion constant. In analogy with using $c$ and $\hbar$ as conversion factors between the meter, second, and kilogram, a relationship that fixes the value of the Boltzmann constant $k$ (where $k = 1.380\ldots \text{x}10^{-23}$ J/K) would be natural and results in the equivalence of 1 K of thermodynamic temperature to $2\pi(138\ 065\ 05/662\ 606\ 93) \text{ x } 10^{11}\ \text{s}^{-1}$. A new definition of the unit kelvin might then be: "The kelvin, unit of thermodynamic temperature, is the change in the thermodynamic temperature of a system whose energy has increased by an amount equal to the energy of a collection of photons whose angular frequencies sum to $2\pi\ (138\ 065\ 05/662\ 606\ 93) \text{ x } 10^{11}$ second$^{-1}$." An equivalent definition that has been proposed in the literature is "The kelvin is the change of thermodynamic temperature that results in a change of thermal energy kT by exactly $1.380\ 650\ 5 \text{ x } 10^{-23}$ joule."[10]

From a microscopic point of view, a further justification for setting $k = 1$ is that the entropy $S = k \ln \Omega$, which is a measure of the number of microscopic states available to a system (which is also a dimensionless number, is now itself a dimensionless quantity.

**B. Current**

The present definition of the unit of current ampere is "The ampere is that constant current which, if maintained in two straight parallel conductors of infinite length, of negligible circular cross-section, and placed 1 meter apart in vacuum, would produce between these conductors a force equal to $2 \text{ x } 10^{-7}$ newton per meter of length."[3] The physical law behind this definition is

$$\frac{F}{l} = \frac{\mu_0}{2\pi}\frac{I_1 I_2}{d} \qquad (1)$$

where $F/l$ is the force per unit length between two parallel infinitely long current-carrying wires, $I_1$ and $I_2$ are the currents in the two wires, and $d$ is the perpendicular distance between the wires. The present SI definition of the ampere thus fixes the value of $\mu_0$ to be $4\pi \text{ x } 10^{-7}$ N/A$^2$ with no uncertainty.

We first re-write this definition in terms of the unit second. In the SI system, force has units of kg•m/s$^2$. Using the conversion factors developed in the previous two sections to express the kilogram and meter in terms of the second, we find that 1 N is equivalent to $2\pi(299\ 792\ 458/662\ 606\ 93) \text{ x } 10^{41}\ \text{s}^{-2}$. In natural units, a force of 1 s$^{-2}$ can be interpreted as the applied force that causes a mass of 1 s$^{-1}$ (roughly $1.173 \text{ x } 10^{-51}$ kg) to accelerate at a rate of 1 s$^{-1}$ (roughly $2.998 \text{ x } 10^{8}$ m/s$^2$). The fact that both mass and acceleration have identical units but are very different quantities should not be any more cause for alarm than the fact that energy and torque have identical units in terms of base SI units.

The definition of ampere can then be rewritten as follows: "The ampere is that constant current which, if maintained in two straight parallel conductors of infinite length, of negligible circular cross-section, and placed 1 second apart in vacuum, would produce between these conductors a force equal to $4\pi(299\,792\,458/662\,606\,93) \times 10^{34}$ second$^{-2}$ per second of length."

Following the example set previously in obtaining equivalences between SI units using physical constants as conversion factors, we derive a numerical equivalence between the ampere and the second by setting $\mu_0$ equal to unity. Since $\mu_0 = 4\pi \times 10^{-7}$ N/A$^2$, we find that 1 ampere is equivalent to $\sqrt{4\pi \times 10^{-7} \times 2\pi \frac{299792458}{66260693} \times 10^{41}}\ s^{-1}$. This numerical equivalence is a matter of convention, however. One could also, on the basis of equation (1), decide to set $\mu_0/2\pi$ equal to unity, so that 1 ampere is equivalent to $\sqrt{\frac{4\pi \times 10^{-7}}{2\pi} \times 2\pi \frac{299792458}{66260693} \times 10^{41}}\ s^{-1}$. Such a definition of ampere makes the magnetic permeability dimensionless.

**C. Mole**

As with the ampere and candela, the mole is already defined in terms of other base SI units. As a unit, the mole is similar to "dozen" in that it represents a specific number of individual objects. The current SI definition is "[1]. The mole is the amount of substance of a system which contains as many elementary entities as there are atoms in 0.012 kilogram of carbon 12; its symbol is 'mol.' [2]. When the mole is used, the elementary entities must be specified and may be atoms, molecules, ions, electrons, other particles, or specified groups of such particles."[3] Similar to the present definition of the kelvin, the definition of the mole is based on a particular substance, in this case, carbon-12.

This definition can be re-written in terms of the unit second as "[1]. The mole is the amount of substance of a system which contains as many elementary entities as there are atoms in an amount of carbon 12 with a rest energy equal to a collection of photons whose angular frequencies sum to exactly $0.012 \bullet 2\pi(299\,792\,458^2/662\,606\,93) \times 10^{41}$ second$^{-1}$; its symbol is 'mol.' [2]. When the mole is used, the elementary entities must be specified and may be atoms, molecules, ions, electrons, other particles, or specified groups of such particles."

Thus, the unit mole remains the same in natural units. The only difference is that the amount of substance in SI units, *i.e.*, 0.012 kilogram of carbon 12, has been expressed in terms of the corresponding angular frequency with the unit second$^{-1}$.

**D. Candela**

The candela, like the other units discussed in this section, is already defined in terms of other SI units. The definition of the SI unit of luminous intensity is "The candela is the luminous intensity, in a given direction, of a source that emits monochromatic radiation of frequency 540 x $10^{12}$ hertz and that has a radiant intensity in that direction of 1/683 watt per steradian."[2] This seemingly odd definition comes from the formula for luminous intensity $I_v$

$$I_v = 683\int_0^\infty I(\lambda)\,\bar{y}(\lambda)\,d\lambda \qquad (2)$$

where $I(\lambda)$ is the radiant intensity of a source (in watt/steradian) and $\bar{y}(\lambda)$ is the standard luminosity function (a dimensionless function with values between 0 and 1 that reflects the sensitivity of the average human eye at different wavelengths; this function peaks with a value of 1 at 555 nm, which is green light with a frequency of 540 x $10^{12}$ hertz). It is interesting to note that while other SI definitions are based on particular substances (such as water) or physical artifacts (the kilogram prototype), the definition of the candela is biologically dependent, based on the response function of an "average" human eye. The purely physical counterpoint to luminous intensity is radiant intensity, which is dimensionally a power per solid angle. Since the steradian is dimensionless in the SI, the candela can be thought of as a special name for watt/steradian in the case of luminous intensity and it is not surprising that in the system of natural units, as in any other system except the SI, luminous (and radiant) intensity have the same dimensions as power.

Converting the unit watt to our base unit of second, using previous conversion factors, we find that 1 watt is equivalent to (2π/662 606 93) x $10^{41}$ $s^{-2}$. A new definition expressed entirely in terms of the base unit second would then read "The candela is the luminous intensity, in a given direction, of a source that emits monochromatic radiation of frequency 540 x $10^{12}$ hertz and that has a radiant intensity in that direction of (1/683) (2π/662 606 93) x $10^{41}$ $s^{-2}$ per steradian." Applying equation (2), we find that 1 candela is equivalent to (1/683) (2π/[299 792 458•662 606 93]) x $10^{41}$ $s^{-2}$ for a monochromatic 540 THz light source.

**E. Other units**

Table I lists some common quantities and their units in both the SI and a natural unit system based on the second. Although in most cases, users of natural units choose energy or length as their fundamental unit, we have used the second because of the precedent set by the definition of the meter in the SI unit system. One can easily convert the table knowing that the second, meter, and inverse kilogram are all dimensionally equivalent in natural units. As mentioned previously, under our choice of constants ($c$, $\hbar$, $k$) to set to unity, a number of quantities, such as velocity, angular momentum, electric charge, entropy, and perhaps somewhat more surprisingly, resistance are dimensionless quantities. Dimensionless velocities are expressed as fractions of the speed of light and angular momenta are in units of the quantum of action $\hbar$ (or $h$ if one so chooses, as discussed in section II.B). Because the electron charge is closely related to the coupling strength of the electromagnetic interaction between an electron and a photon as expressed by the fine structure constant $\alpha$, electric charge is also dimensionless. That the resistance is dimensionless appears to be a coincidence stemming from the fact that both electric potential and current have the same dimensions.

## IV. Other "natural units" systems

Throughout the literature, one can find reference to a number of different systems of "natural units." In these systems, some subset of the constants $c$, $\hbar$, $k$, $G$, $m_e$, $m_p$, $e$, $\varepsilon_0$, and $\mu_0$, are given specified values, usually 1. Perhaps the most well-known of these are Planck units, in which $c = \hbar = k = G = 1$.[11] In the natural units discussed in this paper, we have specified the values of $c$, $\hbar$, $k$, and $\mu_0$.

There is a fundamental difference between the natural unit system that we have discussed in this paper and Planck units. In the system discussed here, "natural" refers to the number of base units. With the exceptions of the mole and candela, which are based on human conventions and physiology and thus whose definitions and equivalences can never be freed of those influences, we have shown that the physical laws of our universe provide a basis for forming equivalences between five of the dimensionally independent base units of the SI unit system and thus that as far as nature is concerned, there is a justification for expressing all quantities using powers of a single base unit. In a sense, physics allows a "unification" of all the units just as physics seeks a unification of the descriptions of all phenomena. Furthermore, when we set the values of the constants $c$, $\hbar$, and $k$ to unity, those constants become dimensionless and the dimensions, as well as the units, of physical quantities are changed.

On the other hand, in natural unit systems such as the Planck units, "natural" refers to the magnitude of the standard for each of the base units. Planck units seek to establish standards that are based on nature as represented by values of the universal constants, rather than based on human convention. So, instead of the meter, which was originally conceived as one ten-millionth of the length of Earth's meridian along a quadrant (essentially one ten-millionth of the distance from the equator to the north pole), as the base standard of length, the standard of length is that formed by a particular combination of universal constants $(hG/c^3)^{1/2}$ that results in a quantity with the dimension of length, which happens to be equivalent to roughly 4 x $10^{-35}$ meter. Similarly, a "natural" unit of mass is not the kilogram (originally conceived of as the mass of a cubic decimeter of water), but a mass equivalent to a combination of universal constants $(hc/G)^{1/2}$ that results in a quantity with the dimension of mass, or 5.5 x $10^{-8}$ kilogram. Thus, in Planck units, quantities are still expressed using multiple base units, although these units are hidden and made implicit by setting $c = \hbar = k =$ $G = 1$. Although the units of physical quantities have been modified, the dimensions of those quantities remain the same, in contrast the system of natural units.

## V. Practical aspects and future definitions of SI base units

From a practical standpoint, there are myriad reasons, ranging from the precision with which experiments can realize the comparison of a given quantity to its defined standard, to the widespread availability of a defined standard, to the uncertainties that are introduced in other quantities based on a given set of definitions of the base SI units, for why the present definitions of the SI units are what they are. For example, it may seem strange that in the twenty-first century, the basis of our unit of mass is a chunk of platinum-iridium alloy cast in 1889 but given the limitations of our technology it is, perhaps incredibly, still the standard that results in the smallest uncertainties for a wide range of other quantities and constants. Likewise, the kelvin is currently defined in terms of a particular state of water, and although variations in the impurity concentration and isotopic composition of a particular sample of water introduce uncertainties in the realization of this standard, those uncertainties are still smaller than those associated with measuring the value of $k$ by other means.

Recently, there have been some discussions concerning revising the definitions of the base SI units so that all of them, or as many as possible, are linked to invariants of nature, rather than physical artifacts.[5,12] This is equivalent to specifying exact values of a number of constants, such as $c$, $h$, $k$, $N_A$, $e$, etc. For example, the current definition of the meter "The meter is the length of the path traveled by light in vacuum during a time interval of 1/299 792 458 of a second." could equivalently be written as "The metre, unit of length, is such that the speed of light in vacuum is exactly 299 792 458 metres per second."[5] and the proposed definition of the kilogram stated earlier "The kilogram is the mass of a body whose equivalent energy is equal to that of a number of photons whose frequencies sum to exactly [(299 792 458)$^2$/662 606 93] x 10$^{41}$ hertz." could equivalently be written "The kilogram, unit of mass, is such that the Planck constant is exactly 6.626 069 3 x 10$^{-34}$ joule second."[5] Likewise, the kelvin, mole, and ampere could be defined by specifying the values of $k$, $N_A$, and $e$, respectively. These definitions would be logically equivalent to the ones we have developed in this paper.

Although such proposals, which have been in the works since 2005, have several advantages and will likely be considered by the 24th CGPM in 2011, some counter-arguments have also been advanced.[13] One counter-argument is that such new definitions might reduce the availability of the standards, since more sophisticated experimental equipment is necessary for making comparisons with microscopic standards than with macroscopic standards. Another is that microscopic standards might benefit only a subset of the community of SI system users, while becoming less accessible or coherent to other groups. Finally, there remains a significant problem to be resolved in the definition of the kilogram in that the various experiments that might be used to determine the value of $h$ are not all in agreement, even within their respective experimental errors. Defining the kilogram by specifying a value of $h$ does not solve this problem, but merely pushes it elsewhere in the determination of other constants. There are those who feel that this inconsistency in the experiments should be resolved first, before redefining the kilogram.

From a theoretical point of view, the precision with which various standards can be measured is irrelevant to the fact that all of the base SI units, and thus all units, can be given physically meaningful definitions in terms of a single fundamental unit, and thus we see that there is indeed a physical basis for the natural unit system. In this paper, we chose the second to be our fundamental unit based on current definitions of SI units, but we could have equally well used any unit and in fact, most practitioners choose energy (mass) as their base unit. We also developed specific numerical equivalences between all of the base SI units, although we have seen that these are not necessarily unique.

Although many particle physicists find the natural unit system well-suited for their work, there are, as we mentioned in the introduction, many practical and pedagogical reasons why it is not convenient for the work of most physicists. From Table I, we can see that the techniques of dimensional analysis or checking the units of an answer lose their usefulness. Just as the judicious choice of a coordinate system can make a seemingly difficult problem easy to solve, it is up to practicing physicists to choose a system of units that best facilitates their work. Furthermore, as pointed out by Levy-LeBlond[14] even after some future time at which technology progresses to the point where all of the SI units can be defined in terms of a single unit, the awkwardness of some of the new definitions, the need to communicate with other disciplines, and plain inertia will likely conspire to preserve the use of multiple units and unit systems (*c.f.*, the continuing widespread use of British units in the US.)

## VI. Status of Fundamental Constants

Now let us consider the status of fundamental constants. From the previous discussion, it is clear that not all of the quantities we call "fundamental physical constants" are alike. Levy-Leblond[14] grouped the fundamental physical constants into two categories, (1) constants characterizing whole classes of physical phenomena (such as the electric charge $e$ and the universal gravitational constant[15] $G$), and (2) universal constants (such as $c$ and $h$) which act as concept or theory synthesizers (for example, Planck's constant $h$ synthesizes the concepts of momentum and wavelength through the relation $p = h/\lambda$).

Based on natural units and dimensional analysis, it is more natural and revealing to group the fundamental constants into two categories, units-independent (A), and units-dependent (B). Because only constants in the former category have values that are not determined by the human convention of units, we argue that they are true fundamental constants in the sense that they are inherent properties of our universe. In comparison, constants in the latter category are not fundamental constants in the sense that their particular values are determined by the human convention of units. These are often historical products of an incomplete physical understanding of our universe. If we were to reformulate physics today from scratch in the simplest possible way using all of our present knowledge, the units-independent constants would still appear as parameters in our theories while the units-dependent constants would not appear at all.[16] Thus, many of the papers in metrology which discuss the measurement of the fundamental physical constants[17,18] actually describe two different types of experiments. One is an experiment that seeks to determine more precisely one of the fundamental numbers that characterize our universe. Another is an experiment that seeks to determine more precisely the conversion factor between the historically independent definitions of two units.

One criterion which can be used to determine the category to which a particular constant belongs is whether or not the value of that constant can be made unity by a suitable and physically based definition of units. This is equivalent to the condition that if the value of a constant can be made exact by a redefinition of units, then that constant is units-dependent and is not a true fundamental constant. For example, quantities such as $c$, $\hbar$, and $k$ are not fundamental constants because they can be made to have an exact value or the value unity by redefining the second, meter, kilogram, and kelvin. The vacuum permittivity $\varepsilon_0$ and permeability $\mu_0$ are likewise units-dependent constants that serve as conversion factors between the SI and CGS systems of units. Although they are not equal to one under present definitions of the units, they could be made unity by a redefinition of the ampere. As a final example, the Stefan-Boltzmann constant $\sigma$ is related to other constants by $\sigma = (2\pi^5 k^4)/(15h^3 c^2)$. In natural units (where $c=\hbar=k=1$), the Stefan-Boltzmann constant has the value $\pi^2/60$ and is exact. Furthermore, by a suitable redefinition of the kelvin (or meter, second, or kilogram), its value could be made unity.

In contrast, the fine structure constant $\alpha$ and the weak mixing angle $\theta_w$ (found in the unified electroweak theory) or the strong coupling constant $\alpha_s$ (in quantum chromodynamics) are fundamental constants since they are dimensionless and thus can not, under any circumstances, be made to have the value one. At present, the only quantities that qualify under this criterion as independent fundamental constants are the coupling constants for two of the three fundamental forces–-$\alpha$ and $\theta_w$ (electroweak), and $\alpha_s$ (strong).

The gravitational constant $G$ is a special and interesting case. Although it plays the role of coupling constant for the gravitational force, because it is not dimensionless (in natural units, it is expressed in units of $s^2$), one might consider the gravitational constant to be units-dependent and hence, not inherent in nature. Until a satisfactory quantum theory of gravity is formulated, it may remain a bit of a mystery as to why $G$ seems not be a fundamental constant by this criterion.[15,19]

In 1963, Dirac made an interesting conjecture regarding fundamental constants in physics, "The physics of the future, of course, cannot have the three quantities $\hbar$, $e$, and $c$ all as fundamental quantities. Only two of them can be fundamental, and the third must be derived from those two. It is almost certain that $c$ will be one of the two fundamental quantities."[20] His reasoning was that "the velocity of light $c$ is so important in the four-dimensional picture, and it plays such a fundamental role in the special theory of relativity, correlating our units of space and time, that it has to be fundamental." However, based on our previous analysis of units, we have seen that neither $\hbar$ nor $c$ is an inherent constant of nature since both are units-dependent. Furthermore, even the value of $e$ can be made unity[5] so it is not a fundamental constant either.

In conclusion, natural units are not merely a calculational convenience, but have a conceptual basis rooted in the nature of our physical universe. For example, the fact that physical laws have the same form in all inertial frames and the three spatial coordinates and time appear in coordinate transformations in a symmetric way is consistent with both distances and time intervals being expressed in the same units. Without this four-dimensional symmetry, there would be no physical basis for defining the meter in terms of the second, although one could certainly do so artificially. Similarly, the dual wave-particle nature of matter provides a physical basis for defining mass in terms of length or time.

Our discussion of the natural unit system leads to a natural classification of constants in physics, units-dependent and units-independent constants, in which only the units-independent constants are truly fundamental in the sense of having values that are inherent characteristics[21] of our universe and not based on human convention. At this point, it appears that the only truly fundamental constants are the coupling constants associated with the electroweak and strong forces.

The work was supported in part by the Jing Shin Research Fund of the University of Massachusetts at Dartmouth. The authors would like to thank the referee for detailed and useful comments in revising the paper.

| Quantity | SI units | natural units |
|---|---|---|
| time | s | s |
| length | m | s |
| mass | kg | $s^{-1}$ |
| electric current | A | $s^{-1}$ |
| temperature | K | $s^{-1}$ |
| luminous intensity | cd | $s^{-2}$ |
| velocity | $m\ s^{-1}$ | dimensionless |
| acceleration | $m\ s^{-2}$ | $s^{-1}$ |
| force | $N = kg\ m\ s^{-2}$ | $s^{-2}$ |
| energy | $J = kg\ m^{2}\ s^{-2}$ | $s^{-1}$ |
| momentum | $m\ kg\ s^{-1}$ | $s^{-1}$ |
| angular momentum | $m^{2}\ kg\ s^{-1}$ | dimensionless |
| electric charge | $C = A\ s$ | dimensionless |
| electric potential | $V = m^{2}\ kg\ s^{-3}\ A^{-1}$ | $s^{-1}$ |
| electric field | $V\ m^{-1} = m\ kg\ s^{-3}\ A^{-1}$ | $s^{-2}$ |
| magnetic field | $A\ m^{-1}$ | $s^{-2}$ |
| resistance | $\Omega = m^{2}\ kg\ s^{-3}\ A^{-2}$ | dimensionless |
| capacitance | $F = m^{-2}\ kg^{-1}\ s^{4}\ A^{2}$ | s |
| inductance | $H = m^{2}\ kg\ s^{-2}\ A^{-2}$ | s |
| entropy | $J\ K^{-1}$ | dimensionless |

Table I. Units of common quantities in the SI system of units and a natural unit system based on the second.